\documentclass[12pt]{article}
\usepackage{graphicx,latexsym}

\def\a{\alpha}
\def\b{\beta}

\def\s{\sigma}

\def\lam{\lambda}
\def\Lam{\Lambda}
\def\gm{\gamma}
\def\Gm{\Gamma}
\def\P{{\rm P}}
\def\gr{{\rm gr}}
\def\QG{{\rm QG}}
\def\sq{\sqrt}
\def\half{\frac{1}{2}}
\def\fr{\frac}

\begin{document}

\begin{titlepage}

\begin{flushright}
{\sc August 2007}
\end{flushright}

\begin{center}
{\Large {\bf Baryogenesis by Quantum Gravity}}
\end{center}

\vspace{5mm}

\begin{center}
{\sc Ken-ji Hamada\footnote{E-mail address: hamada@post.kek.jp}, Azusa Minamizaki\footnote{E-mail address: minamizaki@hep.phys.ocha.ac.jp} and Akio Sugamoto\footnote{E-mail address: sugamoto@phys.ocha.ac.jp}}
\end{center}

\begin{center}
{}$^1${\it Institute of Particle and Nuclear Studies, KEK, Tsukuba 305-0801, Japan} \\
{}$^1${\it Department of Particle and Nuclear Physics, The Graduate University for Advanced Studies (Sokendai), Tsukuba 305-0801, Japan} \\
{}$^{2,3}${\it Department of Physics, Ochanomizu University, Otsuka, Bunkyo-ku, Tokyo 112-8610, Japan}
\end{center}

\begin{abstract}
A novel mechanism of baryogenesis is proposed on the basis of the phase transition from the conformal invariant space-time to the Einstein space-time in quantum gravity. Strong-coupling gravitational excitations with dynamical mass about $10^{17}$GeV are generated at the transition. They eventually decay into ordinary matters. As a realization of unparticle physics we show that the low energy effective interactions between the gravitational potential describing the excitation and the non-conserving matter currents by the axial anomalies can explain matter asymmetry out of thermal equilibrium.

\vspace{5mm}

\end{abstract}
\end{titlepage}

The Wilkinson microwave anisotropy probe (WMAP) experiment \cite{wmap3} has established the inflationary scenario of the universe \cite{guth,starobinsky}. Various cosmological parameters have been determined precisely, in which the baryon number density of the universe has been $n_B/n_\gm = (6.14 \pm 0.25) \times 10^{-10}$. There are many attempts to derive this small number \cite{yoshimura,nw,bsw,krs,fy,ck,hs,dkkms,aps}, but it is known that this baryon excess cannot be produced within the Standard Model of particle physics \cite{hs}. So, it is significant to seek more effective sources of the generation of baryon asymmetry in the early universe. In this paper we propose a new mechanism based on the quantum gravity scenario of the universe without introducing any additional field.

One of the prominent features of quantum gravity scenario we employ here is that there is a dynamical energy scale separating quantum space-time and classical space-time \cite{hy,hhy,hhsy}. In the early epoch of the universe the conformal symmetry has a significant meaning, because there would be a period that space-time is totally fluctuating quantum mechanically so that geometry loses its classical meaning and the scale invariant picture will emerge, while in the present universe there is no such a symmetry. Thus, the existence of the dynamical scale indicates that there was a space-time phase transition in the evolution of the universe according to the violation of the conformal symmetry.

The dynamical scale is introduced in the renormalizable quantum gravity on the basis of a conformal gravity in four dimensions \cite{hamada02}. The inflationary solution requires that the dynamical scale is of the order of $10^{17}$GeV below the Planck scale in order to explain the correct number of e-foldings and the amplitude of cosmic microwave background anisotropies observed by WMAP \cite{hhy,hhsy}. We show that this magnitude of the dynamical scale also has a significant meaning for the generation of the baryon asymmetry.

The basic idea of the mechanism is as follows: scale invariant quantum 
fluctuations\footnote{ 
Physical states are defined satisfying the Wheeler-DeWitt equations of conformal algebra \cite{hh}.
} 
initially covering the whole universe are gradually reduced into the smaller size of fluctuations according to the expansion of the universe. Field fluctuations freeze to localized gravitational objects and eventually decay into ordinary matters. The baryon asymmetry is then generated considering the low energy effective interactions in which those strong-coupling gravitational excitations couple to matter currents. These interactions are acquired provided the currents are not conserved.

This may be a realization of the unparticle physics recently proposed by Georgi \cite{georgi}, since the strong-coupling gravitational excitation considered here has its origin in the breaking of conformal symmetry and the effective interactions introduced are natural ones in the unparticle physics.

\paragraph{Gravitational excitations at strong coupling}
The existence of dynamical energy scale of quantum gravity $\Lam_\QG$ is indicated by the asymptotic freedom of the coupling for the Weyl action $(-1/t^2)C_{\mu\nu\lam\s}^2$, where $C_{\mu\nu\lam\s}$ is the Weyl tensor, in renormalizable quantum theory of conformal gravity with the Einstein action \cite{hamada02}.\footnote{ 
The Wess-Zumino condition requires that the four-derivative actions should be conformal invariant such that the $R^2$ action is forbidden \cite{bcr}, while the Einstein action and the cosmological constant are not forbidden and added. 
} 
The running coupling constant is written as $1/t_r^2(p)=\b_0 \ln(p^2/\Lam_\QG^2)$ for physical momentum $p$ and the beta function was computed as $\b=-\b_0 t_r^3$ with $\b_0 >0$ \cite{ft,hamada02}.

The coupling constant of the Weyl action takes care of the traceless tensor mode in the metric field, which measures a deviation from conformal symmetry. At very high energies, the coupling is vanishing and fluctuations of conformal mode become dominated so that exact conformal invariance is realized, while at the dynamical energy scale the coupling constant diverges and the symmetry is completely broken, turning to the conventional Friedmann universe. Then, energies stored in extra degrees of freedom in higher derivative gravitational fields shift to matter degrees of freedom, causing the big bang. In addition, strong-coupling gravitational objects will be generated. Since the quantum correlation becomes short-ranged about the order of $\xi_\gr=1/\Lam_\QG$, gravitational field fluctuations freeze to localized objects, like glueballs in QCD, with the size of order of $\xi_\gr$ and the mass $m_\gr=\Lam_\QG$. They eventually decay into ordinary matters in the lifetime of order of $\tau_\gr=1/\Lam_\QG$.

In general such a strong-coupling excitation will be described in terms of a composite field, but we here do not care about the detailed structure of the excitation. We simply describe it as a small gravitational fluctuation in the Friedmann background using the gravitational potential, or the Bardeen potential \cite{bardeen}. In the longitudinal gauge, it is defined by $d^2s = a^2 [ -(1-2\Phi)d\eta^2 + (1+2\Phi)d{\bf x}^2 ]$, where $\eta$ and ${\bf x}$ are the comoving coordinates. The scale factor $a$ is the solution of the Friedmann universe satisfying the homogeneous equation of motion $\dot{H} +2 H^2=0$, where $H=\dot{a}/a$ and the dot denotes the derivative with respect to the proper time $\tau$ defined by $d\tau =ad\eta$. The initial value of $H$ at the transition point is given by the order of $\Lam_\QG$.

The amplitude of the gravitational potential generated by the localized excitation is estimated as follows: since $\Phi$ is the Newton's potential, the Poisson equation for $\Phi$ with a source of mass $m_\gr$ leads to $\Phi =G m_\gr/r$. This equation is now effective outside the object $r > \xi_\gr$, while the inside will be described by the strong-coupling dynamics of quantum gravity and thus $\Phi$ would be smoothed without the singularity by the quantum effects. So, the amplitude of the fluctuation with mass $m_\gr$ is roughly evaluated at $r = \xi_\gr$ as
\begin{equation}
    \Phi \sim \fr{G m_\gr}{\xi_\gr} \sim \fr{\Lam_\QG^2}{8\pi M_\P^2}, 
        \label{amplitude}
\end{equation}
where $M_\P=1/\sq{8\pi G}$ is the reduced Planck mass. Since $M_\P$ is larger than the dynamical energy scale $\Lam_\QG$, the amplitude is sufficiently small and thus the use of the Bardeen potential to describe such a gravitational state is justified. The condition $M_\P > \Lam_\QG$ also ensures that the excitation is not a black hole, because the size $\xi_\gr$ is greater than its Schwartzschild radius $Gm_\gr$.

The number density of the strong-coupling objects is estimated as follows: right after the space-time transition such localized fluctuations would appear everywhere densely. So, the number density is roughly one per spatial spherical volume of radius $\xi_\gr$, namely $n_\gr =1/(4\pi \xi_\gr^3/3)=3\Lam_\QG^3/4\pi$.  Thus, the ratio of $n_\gr$ and the photon number density is initially given by the order of unity as
\begin{equation}
      \fr{n_\gr}{n_\gm} \biggr|_{\rm initial} \sim \fr{3\pi}{8\zeta(3)} \sim 1,
\end{equation}
where we use the photon number density $n_\gm = 2\zeta(3)T^3/\pi^2$ evaluated at the temperature $T = \Lam_\QG$ and $\zeta(3)=1.202 \cdots$.

\begin{figure}
\begin{center}
\includegraphics[scale=1]{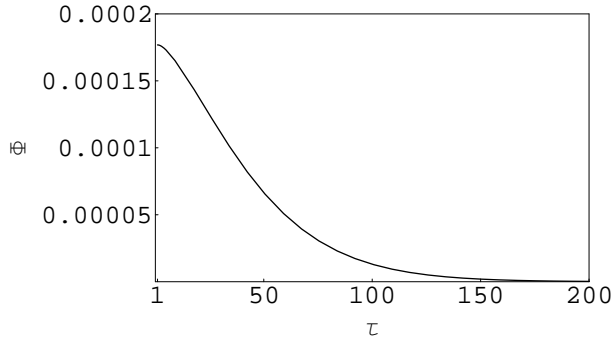}
\end{center}
\caption{\label{fig 1} The evolution of $\Phi$ as a function of $\tau$ for $\Lam_\QG=1/15$ with $M_\P=1$.}
\end{figure}

Since the gravitational objects are uniformly distributed, we remove the spatial dependence of $\Phi$ in the following. The trace of the Einstein tensor leads to the evolution equation $\ddot{\Phi}+5H\dot{\Phi}=0$. We here add the dynamical information that the fluctuation decays in the dynamical time scale. Since they have the mass $m_\gr$ and the decay width $\Gm_\gr$ of order of the mass, the equation would be modified as
\begin{equation}
     \ddot{\Phi} + 5 H \dot{\Phi} +\Gm_\gr \dot{\Phi} +m_\gr^2 \Phi = 0.
     \label{equation}
\end{equation}
The last two terms denote effects of the decay. Since $\Phi$ is a real field and it should monotonically decay and vanish even in the flat space-time of $H=0$, the decay width has to be greater than $2m_\gr$. We here take $\Gm_\gr = 2 m_\gr$, and then $\Phi$ decays exponentially with lifetime $\tau_\gr$ in the flat background. The simulation of equaion (\ref{equation}) is depicted in figure \ref{fig 1}, in which the initial values of $\Phi$ and $H$ are given by (\ref{amplitude}) and $\Lam_\QG$, respectively.

\paragraph{Matter Asymmetry}
We consider the low energy effective theory of quantum gravity which describes the dynamics below $\Lam_\QG$ after the space-time transition was occurred. The key ingredient to produce a matter asymmetry is a C and CP-violating Cohen-Kaplan type interaction \cite{ck,dkkms} between the gravitational potential of strong-coupling excitation $\Phi$ and the non-conserving matter current $J^\mu$:
\begin{equation}
    {\cal L}_{\rm low} = C \fr{\Lam_\QG^2}{M_\P^2} 
            \hbox{$\sq{-\hat{g}}$} \left( \hat{\nabla}_\mu \Phi \right) J^\mu ,
            \label{CK}
\end{equation}
where the cut-off energy scale is taken as $\Lam_\QG$. The coefficient $C$ is a dimensionless constant, which is taken to be the order of unity and the metric $\hat{g}_{\mu\nu}$ denotes the Friedmann background. This type of interaction is recently introduced in \cite{georgi}. The field $\Phi$ corresponds to a kind of unparticle stuff obtained by the violation of conformal invariance at the dynamical scale $\Lam_\QG$. The canonical dimension of the $\Phi$ field is for simplicity set to be zero (it is one if taking the canonical normalization of gravitational fields). Rigorously, it has an anomalous dimension $d_\Phi$ \cite{hy,hh}, and the scale-dependent coefficient should be replaced with $C \Lam_\QG^{2-d_\Phi}/M_\P^2$.

If the thermal equilibrium were realized in the universe right after the space-time transition was occurred, the interaction would dynamically violates CPT invariance leading to $\mu_M n$, where $n=J^0$ is the number density of matter and $\mu_M = C (\Lam_\QG^2/M_\P^2)\dot{\Phi}$ plays a role of the chemical potential to produce matter asymmetry in the thermal universe. However, since $H$ takes a larger value of the order of $\Lam_\QG$ right after the transition, the universe would not be in thermal equilibrium.

We consider the generation of matter asymmetry out of thermal equilibrium. Taking into account the Cohen-Kaplan type interaction, the equation of motion is written as
\begin{equation}
    3 M_\P^2 \left( \ddot{\Phi} + 5H \dot{\Phi} + \Gm_\gr \dot{\Phi} + m_\gr^2 \Phi \right)
    = f_{CK} \left( \dot{n} +3H n \right), 
        \label{equation-CK}
\end{equation}
where $f_{CK}=C \Lam_\QG^2/M_\P^2$.

Among various CP-violating decay processes, we consider those via axial anomalies. Since gauge bosons couple to quarks and leptons left and right asymmetrically as in the Standard Model, the baryon and lepton number currents are violated through the axial anomaly:
\begin{equation}
     \partial_\mu J_{B, L}^\mu = N_g \fr{\alpha}{4\pi} 
         \epsilon^{\mu\nu\lam\s}tr \left( F_{\mu\nu}F_{\lam\s} \right)
         \label{FF}
\end{equation}
where $N_g$ is the number of generations and $\alpha=g^2/4\pi$ is the gauge coupling constant. Thus, the divergence of the current in interaction (\ref{CK}) is replaced with the r.h.s. of (\ref{FF}), which leads to a dimension-five interaction with the coupling constant $f_{CK}N_g(\alpha/4\pi)/\sq{3}M_\P$, where the canonical normalization of kinetic term is taken into account for $\Phi$. Thus, the decay width of the $\Phi$ field through this process is evaluated as
\begin{equation}
    \Gm_{CK} = \left( \fr{N_g f_{CK}}{\sq{3}M_\P} 
                    \fr{\alpha}{4\pi} \right)^2 m_\gr^3 
\end{equation}
where $m_\gr$ is the mass of the decaying gravitational object.

Incorporating the decay process with $\Gm_{CK}$ into the equation of motion of $\Phi$, we obtain 
\begin{equation}
     \ddot{\Phi} + 5H \dot{\Phi} + \Gm_\gr \dot{\Phi} 
          + m_\gr^2 \Phi +\Gm_{CK} \dot{\Phi}=0.
     \label{equation-Gm}
\end{equation}
Combining equations (\ref{equation-CK}) and (\ref{equation-Gm}) and identifying $n$ with $n_{B,L}$, we obtain the equations for the baryon and lepton number densities
\begin{equation}
    \dot{n}_{B, L} +3H n_{B, L} 
    = -\fr{1}{f_{CK}} 3M_\P^2 \Gm_{CK} \dot{\Phi}
    = -C N_g^2 \left( \fr{\alpha}{4\pi} \right)^2 
              \fr{\Lam_\QG^5}{M_\P^2} \dot{\Phi} .
       \label{equation-n}
\end{equation}
Solving the coupled equations of (\ref{equation-Gm}) and (\ref{equation-n}), we can obtain the number densities.

Before carrying out the numerical simulation, we give a rough estimation of the ratio $Y_{B,L}=n_{B,L}/n_\gm$. Integrating equation (\ref{equation-n}) we obtain the following expression:
\begin{equation}
    Y_{B,L} = - C N_g^2 \left( \fr{\alpha}{4\pi} \right)^2 
              \fr{\Lam_\QG^5}{M_\P^2}
          \int^\infty_{\tau_0} d\tau  \fr{\dot{\Phi}}{n_\gm}
      \sim  C N_g^2 \left( \fr{\alpha}{4\pi} \right)^2 
              \fr{\Lam_\QG^5}{M_\P^2} 
           \fr{\Phi}{n_\gm} \biggr|_{\tau_0}, 
\end{equation}
where $\tau_0$ is the initial time when the transition was occurred. In the second line we take $n_\gm$ a constant for simplicity and use the fact that $\Phi$ finally vanishes. Substituting the initial amplitude of $\Phi$ given by (\ref{amplitude}) and the photon density $n_\gm \simeq 2\zeta(3)\Lam_\QG^3/\pi^2$, we obtain
\begin{eqnarray}
    Y_{B,L} &\sim& C \fr{\pi}{16\zeta(3)} N_g^2 \left( \fr{\alpha}{4\pi} \right)^2
           \fr{\Lam_\QG^4}{M_\P^4}.
\end{eqnarray}
Here, the reduced Planck mass is $M_\P=2.4 \times 10^{18}$GeV and the number of generation is $N_g=3$. The strength of coupling constant is taken to be the unification value of $\alpha =1/45$ as popularly accepted. If we take the coefficient and the ratio of two mass scales as $C=1$ and $\Lam_\QG/M_\P \sim 1/10$, respectively, we obtain the magnitude of order of $10^{-10}$.

\begin{figure}
\begin{center}
\includegraphics[scale=1]{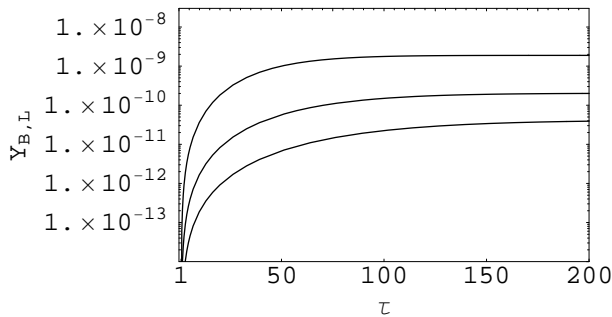}
\end{center}
\caption{\label{fig 2}  Numerical simulations of $Y_{B,L}$ for $\Lam_\QG/M_\P=1/10$(top), $1/15$(middle), and $1/20$(bottom).}
\end{figure}

Numerical results for various $\Lam_\QG$ are depicted in figure \ref{fig 2}. The simulations are carried out with the initial conditions: $\Phi(\tau_0)=\Lam_\QG^2/8\pi M_\P^2$, $Y_{B,L}(\tau_0)=0$, $H(\tau_0)=\Lam_\QG$ and the initial time is set to be $\tau_0=1$. The temperature determining the photon number density is defined by $T=(30\rho/\pi^2 g_*)^{1/4}=(90/\pi^2 g_*)^{1/4} \sq{M_\P H}$, where $\rho$ and $g_*$ are the energy density and the effective number of massless degrees of freedom, respectively. $g_*$ is taken to be $100$. 

Finally, we discuss the decay process through the gravitational axial anomaly. An advantage of this process is that we do not need to care about the washout of $B+L$ by the sphaleron process \cite{krs} in the late universe. Rather, the sphaleron process is here used to generate the baryon number from the lepton number initially generated \cite{fy}. Recently, an idea of leptogenesis through the gravitational anomaly in the inflationary background was proposed in \cite{aps}, while our approach is the alternative one generating lepton asymmetry right after the big bang.

Since the gravitational field couples with the left- and right-handed matter fields equally, the anomaly requires an imbalance of left- and right-handed fermions. So, we assume that there is no right-handed neutrino, and thus we obtain
\begin{equation}
     \partial_\mu J_L^\mu = \fr{N_\nu}{16\pi^2} \half \epsilon^{\mu\nu\lam\s}
         C_{\mu\nu\a\b} C_{\lam\s}^{~~~\a\b} ,
\end{equation}
where $N_\nu=3$ is the number of species of left-handed neutrinos. Thus, we can produce the lepton asymmetry through the gravitational anomaly and the baryon asymmetry can also be generated, provided the sphaleron process inter-converting baryons and leptons are effective.

The decay from the localized gravitational object to gravitons through gravitational anomaly is caused by the dimension-seven interaction, and thus the decay width is estimated as
\begin{equation}
    \Gm^\gr_{CK} \sim \left( \fr{f_{CK}}{\sq{3}M_\P} 
                    \fr{N_\nu}{16\pi^2 M_\P^2} \right)^2 m_\gr^7 .
\end{equation}
As in the same way discussed above, we obtain the equation for the lepton number density,
\begin{equation}
    \dot{n}_L +3H n_L 
    = -\fr{1}{f_{CK}} 3M_\P^2 \Gm^\gr_{CK} \dot{\Phi}.
\end{equation}
Thus, the lepton-to-photon ratio is estimated as
\begin{eqnarray}
    Y_L &\sim& C \fr{N_\nu^2}{16^3\pi^3\zeta(3)}
           \fr{\Lam_\QG^8}{M_\P^8}.
\end{eqnarray}
This value is rather smaller than that obtained using the gauge axial anomaly, though we could get the observed magnitude of asymmetry by taking the difference of two mass scales small.

\paragraph{Conclusion}
We have presented a new mechanism that generates the cosmic baryon asymmetry on the basis of the space-time phase transition of quantum gravity origin. The dynamical energy scale of the space-time transition is given by the order of $10^{17}$GeV below the Planck scale. This scale also characterizes the localized strong-coupling gravitational excitation of quantum gravity. We have studied the CP-violating low energy effective interaction between the localized gravitational object and non-conserving baryon and lepton number currents. It has been shown that the correct magnitude of the baryon asymmetry can be generated through the axial anomalies out of thermal equilibrium.


\end{document}